\pdfoutput=1
\documentclass[journal]{IEEEtran}

\usepackage{booktabs}
\usepackage{graphicx}
\usepackage{cite}
\usepackage{amsmath,amssymb,amsfonts}
\usepackage{url}
\usepackage{xcolor}
\usepackage[hidelinks]{hyperref}
\hypersetup{pdftitle={Comb Test Under a Non-Uniform Null: Centered Total Variation and Phase-Blindness},pdfauthor={Nikola Banić and Neven Elezović}}
\usepackage[normalem]{ulem}
\usepackage{stfloats}
\usepackage[final,expansion=true,protrusion=true,tracking=false]{microtype}

\graphicspath{{./}}

\newtheorem{proposition}{Proposition}
\newtheorem{corollary}{Corollary}

\begin{document}

\title{Comb Test Under a Non-Uniform Null: Centered Total Variation and Phase-Blindness}

\author{Nikola Bani{\'{c}} and Neven Elezovi{\'{c}}
}

\markboth{IEEE Signal Processing Letters}
{Bani{\'{c}} and Elezovi{\'{c}}: Comb Test Under a Non-Uniform Null}
\maketitle

\renewcommand\footnoterule{{\hrule height 0pt}}
\let\thefootnote\relax\footnotetext{
Manuscript received. Both authors contributed equally to this work.

Nikola Bani{\'{c}} is with Gideon Brothers, 10000 Zagreb, Croatia (e-mail: nbanic@gmail.com). Neven Elezovi{\'{c}} is with the Department of Applied Mathematics, Faculty of Electrical Engineering and Computing, University of Zagreb, 10000 Zagreb, Croatia (e-mail: neven.elezovic@fer.hr).
}

\begin{abstract}
The comb test detects alternating histogram defects, every second bin raised. For a non-uniform null it is defined on counts centered on their expected values, keeping it blind to which bins are raised; uncentered counts would expose only one choice of raised bins. Its exact null is computed for any rational null, a first beyond the uniform null for the total variation of counts, and an index known before any data predicts how much uncentered counts would lose. It applies where the null is known: converter code densities, bit-source run lengths, image histograms. With real image histograms as nulls it has about twice the power of pooled chi-square at a 6\% mismatch, and it detects a two-lane defect in simulated bit-source run lengths. Code and data are available at~\url{https://github.com/DiscreteTotalVariation/ResidualCombTest}.
\end{abstract}


\IEEEpeerreviewmaketitle

\newcommand{\E}[1]{\mathbb{E}\left[ #1 \right]}
\newcommand{\A}[1]{\lvert #1 \rvert}
\newcommand{\Var}[1]{\mathrm{Var}\left[ #1 \right]}
\newcommand{\V}[1]{\left\lVert #1 \right\rVert_{V}}
\newcommand{\Norm}[2]{\left\lVert #1 \right\rVert_{#2}}
\newcommand{\sgn}{\operatorname{sign}}
\newcommand{\dr}{d_{\mathbf{r}}}
\newcommand{\du}{d_{\mathbf{u}}}

\section{Introduction}
\label{sec:introduction}

\IEEEPARstart{L}{et} $\mathbf{x}=(x_1,\ldots,x_n)\sim\mathrm{Mult}(N,\mathbf{p})$ be a histogram with $n$ ordered bins, $N=\sum_i x_i$ counts, and bin probabilities $\mathbf{p}$, tested against the null hypothesis $H_0$: $\mathbf{p}=\mathbf{r}$ for a fully specified $\mathbf{r}=(r_1,\ldots,r_n)$. The comb test (CT)~\cite{banic2026comb} tests the \emph{uniform} null $\mathbf{r}=\mathbf{u}=(1/n,\ldots,1/n)$ through the discrete total variation (DTV)~\cite{banic2020tvor}
\begin{equation}
    \label{eq:dtv}
    \du(\mathbf{x})=\V{\mathbf{x}}=\sum_{i=1}^{n-1}\A{x_{i+1}-x_i}
\end{equation}
of the counts, with an exact null by dynamic programming. Its value lies in one property: its advantage over chi-square is concentrated on period-2 departures $p_i=1/n+(-1)^{i+1}\delta$, \emph{combs}, such as the alternating code widths of a converter with a mismatched least-significant capacitor~\cite{razavi2017principles}, and needs no knowledge of which parity of bins is raised. This property is called \emph{phase-blindness} here. The letter~\cite{banic2026comb} lists as future work ``a non-uniform null via multinomial weights''; many ordered histograms have no uniform null: converter codes under a sine are arcsine, bit-source run lengths geometric, and an image's code histogram can serve as the null for its requantization.

Exact nulls under a non-uniform multinomial have been computed, without enumerating the sample space, only for statistics of restricted form. Filina and Zubkov~\cite{filina2008exact} call a statistic \emph{decomposable} if it has the form $\sum_i f_i(x_i)$, and compute Pearson's chi-square $X^2$ exactly by a Markov chain over the bins, after~\cite{zubkov2002computational}; that class is treated in~\cite{hirji1997comparison,selivanov2006calculation,baglivo1992methods,keich2006fast,resin2023simple,dwyer2026cellseparable}. Corrado~\cite{corrado2011exact} handles the maximum, minimum, and range of the counts, not decomposable but order-invariant (unchanged under bin permutation). Zubkov and Filina~\cite{zubkov2020computation} give a general chain for statistics updatable from a carried state, instanced on the partial sums (discrete Kolmogorov--Smirnov, equiprobable), on which the discrete Cram\'{e}r--von Mises and Anderson--Darling statistics~\cite{choulakian1994cramer} are also built. Kuriki et al.~\cite{kuriki2018multiplicity} compute exact $p$-values of scan statistics, functions of windows of consecutive cells, through a Markov property. The DTV depends on \emph{adjacent} counts $(x_i,x_{i+1})$ jointly, so it is neither decomposable, order-invariant, nor accumulable from the running partial sum alone; no tractable null of it beyond the uniform case~\cite{banic2026comb} has been published.

For a non-uniform null $\mathbf{r}$ this letter defines the comb test through the DTV of the counts centered on their expected values, the \emph{centered total variation} (TVC) $\dr$, which for $\mathbf{r}=\mathbf{u}$ is~\eqref{eq:dtv}. This letter (i) computes the exact null of the CT for any rational $\mathbf{r}$, for few bins at a small multiple of the uniform case's cost by a residue argument, to the authors' knowledge a first for the DTV under a non-uniform multinomial; (ii) shows what the uncentered $\du$ would measure instead: once the expected difference between neighboring bins exceeds its noise, $\du$ collapses to a fixed signed sum of the counts that a comb raises for one choice of raised bins and lowers for the other, losing phase-blindness (Proposition~\ref{prop:locking}); (iii) gives the collapse index, computable from the null and $N$ before any data, which says how much of that property $\du$ would keep; and (iv) shows on a simulated two-lane bit source and real image histograms where the CT helps and where other tests are preferable. Results use Monte Carlo (MC) or the gamma (Section~\ref{subsec:fitting}) unless marked exact.

\section{The comb test with null $\mathbf{r}$ and its exact null}
\label{sec:exact}

For a null $\mathbf{r}$ the CT applies the DTV to the centered counts $\mathbf{e}=\mathbf{x}-N\mathbf{r}$,
\begin{equation}
    \label{eq:rct}
    \dr(\mathbf{x})=\V{\mathbf{e}}=\sum_{i=1}^{n-1}\A{(x_{i+1}-x_i)-N(r_{i+1}-r_i)},
\end{equation}
which for $\mathbf{r}=\mathbf{u}$ is~\eqref{eq:dtv}. Over real $\mathbf{x}$ with $\sum_i x_i=N$ its minimum is $0$, attained only at $\mathbf{x}=N\mathbf{r}$, so $\dr$ at any alternative's expected histogram is positive and an upper-tail (one-sided) test is natural. The two-sided variant announced in~\cite{banic2026comb} is a different, open question: the uniform null against too-smooth histograms.

\subsection{Calculation specifics}
\label{subsec:specifics}

Rational probabilities $r_i=q_i/L$, positive integer weights $\mathbf{q}=(q_1,\ldots,q_n)$, $L=\sum_i q_i$, are assumed; this covers dither, run-length, and synthesized nulls exactly and any other to a precision paid for in cost (below); and
\begin{equation}
    \label{eq:S}
    S(\mathbf{x})=L\,\dr(\mathbf{x})=\sum_{i=1}^{n-1}\A{L(x_{i+1}-x_i)-N(q_{i+1}-q_i)}
\end{equation}
is an integer: centering introduces the denominator $L$ that the weights already supply, so the device used for $X^2$ in~\cite{good1983exact,filina2008exact} carries over. $\dr=S/L$ is thus \emph{lattice-valued}, its values multiples of the step $1/L$, so they index integer states of a recurrence. Let $C(i,M,m,s)$, $0\le m\le M\le N$, $0\le s\le 2NL+N\sum_{i=1}^{n-1}\A{q_{i+1}-q_i}$, be the total weight of placements of $M$ of the $N$ observations into the first $i$ bins with $m$ in bin $i$ and statistic $s$ so far, each observation in bin $j$ weighted by $q_j=Lr_j$, so that $\prod_j q_j^{x_j}=L^N\prod_j r_j^{x_j}$ and the total divided by $L^N$ is a probability; such weighted schemes are developed in full in~\cite{filina2008exact,zubkov2020computation}. For $i=1$ the only state has $m=M$ and $s=0$: $C(1,M,M,0)=\binom{N}{M}q_1^M$, and $C(1,M,m,s)=0$ otherwise. For $i>1$, with $k$ counts in bin $i-1$, appending $m$ observations to bin $i$ adds $\A{L(m-k)-N(q_i-q_{i-1})}$ to the statistic and multiplies the weight by $\binom{N-M+m}{m}q_i^m$; summing over all $k$ from $0$ to $M-m$ gives
\begin{equation}
    \label{eq:dp}
    \begin{gathered}
    C(i,M,m,s)=\binom{N-M+m}{m}q_i^m\\
    \cdot\sum_{k=0}^{M-m}C\bigl(i-1,M-m,k,s-\A{L(m-k)-N(q_i-q_{i-1})}\bigr),
    \end{gathered}
\end{equation}
with $C=0$ outside these ranges, so the single-bin case is the only initialization, and $\Pr_{H_0}[S=s]=\sum_m C(n,N,m,s)/L^N$. Equation~\eqref{eq:dp} is the recurrence of~\cite[eq.~(4)]{banic2026comb} with the increment $\A{m-k}$ replaced and the placement factor weighted by $q_i^m$. It is also an instance of the Markov-chain scheme of~\cite[Sec.~4]{zubkov2020computation}, carrying $(s,\text{previous count})$, akin to the finite Markov-chain imbedding of~\cite{fu1994distribution} and to the scan-statistic recursion of~\cite{kuriki2018multiplicity}.

\emph{Cost.} Every term of~\eqref{eq:S} is congruent to $\pm N(q_{i+1}-q_i)$ $\bmod L$, so the reachable $S$ values lie in $c\le\min(L,2^{n-1})$ residue classes $\bmod L$ and number $O(cN)$, not $O(LN)$, with $c$ set by $\mathbf{q}$ and $N$ ($1$ to $16$ here, $N\le100$). Restricted to these, the cost is that of~\cite{banic2026comb}, $O(nN^4)$ time and $O(N^3)$ memory, times a factor set by $c$, not $L$; on the run-length null of Section~\ref{subsubsec:runs} the run time is $2.5$, $2.0$, and $1.3$ times that of the same program on $\du$ at $N=50$, $100$, $200$, as $c$ falls from $8$ to $2$. With many bins $c$ approaches $L$, and the gamma of Section~\ref{subsec:fitting} is used.

\emph{Test.} At level $\alpha$ the CT rejects when $S$ exceeds the smallest $s$ with $\Pr_{H_0}[S>s]\le\alpha$; the $p$-value of $s$ is $\Pr_{H_0}[S\ge s]$. Power under an alternative $\mathbf{p}=\mathbf{q}'/L'$ is exact by the same recurrence with weights $q'_i$ and the null's centering kept. For $\mathbf{q}=(1,\ldots,1)$, i.e., $\mathbf{r}=\mathbf{u}$, \eqref{eq:S} gives $S=n\du(\mathbf{x})$, recovering the distribution of~\cite{banic2026comb}.

\section{What the uncentered variation would measure}
\label{sec:degeneracy}

The statistic~\eqref{eq:rct} is defined with the null built in. Were the uncentered $\du$ used on a non-uniform null instead, it would not be a comb test: the expected slopes would add a deterministic staircase of order $N$ to~\eqref{eq:dtv}, and beyond losing power it would detect one comb sign and \emph{anti-detect} the other, rejecting it less often than under $H_0$, the more so as $N$ grows. This section gives the mechanism behind the $\du$ columns reported below. Let $g_i=p_{i+1}-p_i$ be the expected slope between bins $i$ and $i+1$, and $D_i=x_{i+1}-x_i$ its sample counterpart, so $\E{D_i}=Ng_i$.

\begin{proposition}[sign locking]
\label{prop:locking}
For $\mathbf{x}\sim\mathrm{Mult}(N,\mathbf{p})$, null or alternative, with every $g_i\neq0$, $\Pr[\sgn D_i\neq\sgn g_i]\le\exp(-Ng_i^2/2)$, so $E=\bigcap_{i=1}^{n-1}\{\sgn D_i=\sgn g_i\}$ has probability at least $1-\sum_{i=1}^{n-1}\exp(-Ng_i^2/2)$; on $E$,
\begin{equation}
    \label{eq:linear}
    \du(\mathbf{x})=\sum_{i=1}^{n-1}\sgn(g_i)\,D_i=:\Lambda(\mathbf{x}),
\end{equation}
a linear functional of the counts, $\pm(x_n-x_1)$ for monotone $\mathbf{p}$; the null's $\Lambda$ locks under alternatives that keep every $\sgn g_i$.
\end{proposition}
\begin{IEEEproof}
$D_i$ is a sum of terms in $\{-1,0,1\}$ over the $N$ observations (one for bin $i+1$, minus one for bin $i$, zero for any other bin), independent with mean $g_i$; Hoeffding's inequality~\cite{hoeffding1963probability} for $N$ terms of range $2$ gives $\Pr[D_i-Ng_i\le-t]\le\exp(-2t^2/(4N))$ (likewise for the upper tail), and $t=N\A{g_i}$ is the deviation at which the sign flips. On $E$, $\sgn D_i=\sgn g_i$, so $\A{D_i}=\sgn(g_i)D_i$ for every term of~\eqref{eq:dtv}, which is then the signed sum~\eqref{eq:linear}; for monotone $\mathbf{p}$ all signs agree, the interior counts cancel in pairs, and only $\pm(x_n-x_1)$ survives.
\end{IEEEproof}

\begin{corollary}[loss of phase-blindness]
\label{cor:phase}
Let $\mathbf{p}_\pm=\mathbf{r}\pm\boldsymbol\delta$ with $\sum_i\delta_i=0$, and $\Lambda$ as in~\eqref{eq:linear} for the null's slopes. Then $\E{\Lambda(\mathbf{x})\,|\,\mathbf{p}_\pm}-\E{\Lambda(\mathbf{x})\,|\,\mathbf{r}}=\pm N\Lambda(\boldsymbol\delta)$ exactly: the two \emph{comb signs} $\mathbf{p}_\pm$ shift $\Lambda$ oppositely. The rejection rate of an upper-tail rule $\Lambda>\kappa$ is a polynomial in the amplitude of $\boldsymbol\delta$ whose derivative at zero, $\mathrm{Cov}_{H_0}(1\{\Lambda>\kappa\},\sum_j\delta_jx_j/r_j)$, changes sign with the comb sign, so unless it vanishes the rule is to first order above its \emph{size}, its rejection rate under $H_0$, against one sign and below it against the other. Also $\du(\mathbf{x})\ge\Lambda(\mathbf{x})$, with equality on $E$, so under a $\mathbf{p}$ with the null's slope signs the rates of $\du>\kappa$ and $\Lambda>\kappa$ differ by at most $\Pr[E^c]$.
\end{corollary}

\begin{figure*}[!t]
    \centering
    \includegraphics[width=\textwidth]{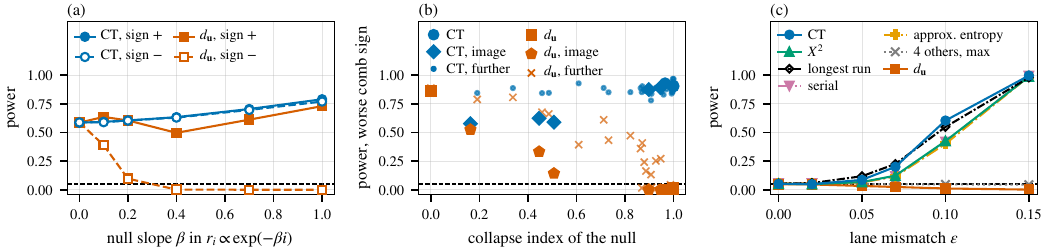}
    \caption{(ii), (iii), (iv); power at $\alpha=0.05$, higher is better, dashed: $\alpha$. a)~What centering changes: against a multiplicative period-2 comb of either sign ($+$: odd bins raised) at $\ell_1$ distance ($\sum_i\A{\delta_i}$) $0.08$ from the null $r_i\propto\exp(-\beta i)$, $n=6$, $N=1000$, the two powers of $\du$ separate as the null steepens, the CT's do not. b)~The collapse index predicts it: worse-sign power versus the null's index, $N=2000$, comb as in (a), for five synthetic nulls, five image nulls ($n=64$), and forty-two further nulls. c)~The two-lane bit source, $2000$-bit windows ($n=6$, $N\approx1000$ runs): serial and approximate entropy at their best block length $m$; ``4 others, max'': best of frequency, runs, spectral, and serial at $m=2$; the starting lane is fixed, favoring the longest-run test.}
    \label{fig:main}
\end{figure*}

Centering applies to the \emph{counts}: subtracting a constant from $\V{\mathbf{x}}$, as the TVOR method~\cite{banic2020tvor} did, leaves~\eqref{eq:linear} a signed sum. The centered counts carry no staircase: under $\mathbf{p}_\pm$, $\mathbf{e}=\pm N\boldsymbol\delta+\boldsymbol\xi$, where $\boldsymbol\xi$ is zero-mean sampling noise whose law differs between the signs only at order $\boldsymbol\delta$, and $\V{\mathbf{e}}$ is even in $\mathbf{e}$, so the two powers of the CT differ only through the noise asymmetry: first order in $\boldsymbol\delta$, but with a derivative at most of order $N^{-1/2}$ times that of $\Lambda$ (repository). Fig.~\ref{fig:main}a shows the CT and $\du$ on $r_i\propto\exp(-\beta i)$: as $\beta$ grows from $0$ to $1$, $\du$ reaches $0.730$ against one sign and $0.000$ against the other, while the CT's two powers stay within $0.02$ of each other, the worse rising from $0.586$ to $0.770$. Two-sided rejection on $\du$ would leave the two signs unequal and lose power on the flat null (repository).

\emph{Collapse index.} Under $H_0$ the bound of Proposition~\ref{prop:locking} is informative once $N\min_i g_i^2$ exceeds $2\log(n-1)$, as on steep nulls of few bins, and vacuous wherever some $g_i$ is near zero, as on every image null here; the collapse is therefore measured. The loss depends not on ``non-uniform'' as such but on how far~\eqref{eq:dtv} has collapsed into~\eqref{eq:linear}, measured by the $H_0$ correlation of $\du(\mathbf{x})$ with $\Lambda(\mathbf{x})$ ($0$ if $\Lambda\equiv0$); this \emph{collapse index} needs only $\mathbf{r}$, $N$, and null draws ($4\cdot10^4$ here). Fig.~\ref{fig:main}b plots the worse-sign power of the CT and of $\du$ against it at $N=2000$ for fifty-two nulls, five of them image nulls. The index \emph{orders} the loss rather than cutting it off: the worse-sign power of $\du$ declines with it (Spearman $-0.90$ to $-0.84$ over five seeds of the further nulls), from $0.85$--$0.89$ at index $0$ to $0.43$ near $0.75$ and to at most $\alpha$ at the top, while the CT never falls below $0.57$.

\section{Calibration beyond the exact regime}
\label{sec:approximate}

\subsection{Gamma approximation}
\label{subsec:fitting}

Beyond the exact regime, the letter~\cite{banic2026comb} prefers a gamma to the better-fitting beta as more conservative in the upper tail, as here. Over five weight families, $N=10$ to $160$, the gamma cut attains exact size $0.039$ to $0.054$. Against the exact cut the gamma is conservative at smaller levels: over four nulls at $N=40$ to $100$ its cut lies above the exact one, with exact sizes $0.007$ to $0.009$ at $\alpha=0.01$ and $0.00035$ to $0.0006$ at $\alpha=0.001$. As in~\cite[Sec.~IV-C3]{banic2026comb}, parameters are moment-matched to $5\cdot10^4$ null draws; on an independent batch the sizes of $\dr$ and the $T(t)$ of Section~\ref{subsec:family} at $\alpha=0.05$ are $0.047$ to $0.052$ for five nulls (one with $n=64$) up to $N=5\cdot10^4$.

\subsection{The family $T(t)$ and the min-$p$ combination}
\label{subsec:family}

Weighting the centered counts bin-wise, an item listed as future work in~\cite{banic2026comb}, gives a family. Let $y_i=e_i/(Nr_i(1-r_i))^{t/2}$ and $T(t)=\sum_{i=1}^{n-1}\A{y_{i+1}-y_i}$, so $T(0)=\dr$, the default, and $T(1)$ is the DTV of the adjusted residuals~\cite{haberman1973residuals}. As in the power-divergence family~\cite{cressie1984multinomial}, none is uniformly best: against a comb, \emph{multiplicative} probabilities $r_i(1+(-1)^{i+1}\varepsilon)$ favor $T(0)$, as in~\cite{perkins2014deficiencies}; \emph{additive} ones $r_i+(-1)^{i+1}\delta$ favor $T(1)$, the gap reaching $0.22$; other shapes keep no fixed order. A practitioner cannot know which holds; hence Tippett's combination~\cite{tippett1931methods}, \emph{min-$p$}: the minimum of the empirical $p$-values of $T(0)$, $T(0.5)$, and $T(1)$ from one null batch, not itself a $p$-value and so cut at the $\alpha$-quantile of an independent second batch~\cite{westfall1993resampling}. Over four nulls and both comb kinds at $N=500$ its worst shortfall behind the per-case winner is $0.03$, against $0.08$ for the best member fixed beforehand, $0.22$ for $T(0)$, and $0.19$ for $X^2$. Only $T(0)$ has the exact null of Section~\ref{sec:exact}.

\section{Experimental results}
\label{sec:results}

\begin{table}[!t]
\centering
\caption{(i), (ii): Exact size and power at $N=40$, $\alpha=0.05$}
\label{tab:exact}
\scriptsize
\setlength{\tabcolsep}{3.2pt}\renewcommand{\arraystretch}{0.92}
\begin{tabular}{@{}lc cc cc@{}}
\toprule
 & & \multicolumn{2}{c}{CT} & \multicolumn{2}{c}{$d_{\mathbf u}$} \\
\cmidrule(lr){3-4}\cmidrule(lr){5-6}
null & index & size & power $+$\,/\,$-$ & size & power $+$\,/\,$-$ \\
\midrule
flat & 0.00 & 0.042 & 0.172\,/\,0.196 & 0.042 & 0.172\,/\,0.196 \\
ramp & 0.29 & 0.042 & 0.176\,/\,0.201 & 0.049 & 0.195\,/\,0.192 \\
tent & 0.50 & 0.042 & 0.166\,/\,0.170 & 0.045 & 0.080\,/\,0.243 \\
run-length & 0.72 & 0.045 & 0.183\,/\,0.230 & 0.047 & 0.252\,/\,0.041$^{\dagger}$ \\
comb & 0.96 & 0.050 & 0.190\,/\,0.176 & 0.047 & 0.002$^{\dagger}$\,/\,0.332 \\
\bottomrule
\end{tabular}

\vspace{2pt}\parbox{\columnwidth}{\scriptsize $+$\,/\,$-$: comb sign of $\mathbf{r}\pm\boldsymbol\delta$ ($+$ raises odd bins), a multiplicative period-2 comb; higher power is better; size: rejection rate attained under $H_0$, not $\alpha$. CT: comb test,~\eqref{eq:rct}; $\du$: uncentered~\eqref{eq:dtv}, one-sided. Nulls: flat $\mathbf{q}=(1,1,1,1,1)$ (CT $=\du$ by~\eqref{eq:S}); ramp $(1,2,3,4,5)$; tent $(1,2,3,4,3,2,1)$; run-length $(16,8,4,2,1,1)$; comb $(1,2,1,2,1)$. Index: exact collapse index at $N=40$ ($0$ on flat). $\dagger$: power below size.}
\end{table}

\subsection{Exact size and power}
\label{subsec:exactpower}

Table~\ref{tab:exact} reports exact sizes and powers (no MC) at $N=40$, $\alpha=0.05$, ordered by index. The two comb signs share $\boldsymbol\delta$, so their $\ell_1$ distances $\sum_i\A{\delta_i}$ are equal, $0.196$ to $0.200$. On the flat null the CT and $\du$ coincide, as~\eqref{eq:S} requires; on the ramp, index $0.29$, $\du$ still treats both signs alike and slightly exceeds the CT on the worse sign at a larger size. Down the column its worse-sign power falls to $0.080$, $0.041$, and $0.002$, the last two \emph{below} its own size, while the CT's stays at $0.166$ to $0.183$; on the comb null one comb deepens the zigzag, the other partly cancels it. Neither statistic is unbiased (power below size), as~\cite{good1984bias} found for $X^2$. Over four complete families ($3$-bin weights to $9$ down to $6$-bin weights to $3$) and nine comb amplitudes, $247\,608$ exact cases, power falls below size in $38\%$ of them for $\du$ and $17\%$ for the CT; from three to six bins the CT's worst shortfall falls from $0.045$ (power $0.002$) to $0.004$; that of $\du$ is its full size, power below $0.001$.

\subsection{Where the CT is and is not sensitive}
\label{subsec:where}

Table~\ref{tab:shapes} fixes the $\ell_1$ distance and varies the null and comb kind. Besides $X^2$ it includes two order-aware competitors of~\cite{banic2026comb}, on $\mathbf{e}$: alternating contrast (AC) and von Neumann ratio (vN)~\cite{von1941distribution}, and two ordinal ones, Cram\'{e}r--von Mises (CvM) and Anderson--Darling (AD)~\cite{choulakian1994cramer}. First, on the comb rows the best test other than AC, which is tied to period~2, is the CT, $T(1)$, or min-$p$; a Neyman smooth test~\cite{neyman1937smooth} on two to four frequencies (one is AC) trails the CT on four of five nulls (repository). The margin over $X^2$ depends on geometry: for a multiplicative comb on the run-length null $0.573$ against $0.382$, for an additive comb $T(1)$ and min-$p$ lead, $0.842$ and $0.816$ against $0.806$; in Block~1 min-$p$ stays within $0.03$ of the better member. Second, CvM and AD are weak on combs ($0.06$ to $0.34$) but, like $X^2$, strong on linear tilts ($0.58$ to $0.88$), where the CT is weak (repository): the families are complementary.

Block 2 follows~\cite[Sec.~IV-C3]{banic2026comb} without its rank transform: the null is the clean image's exact code histogram~\cite{vandewalle2014scikit}, and $N$ requantized pixels are tested. Over five images (three shown; all ten usable images: $1.63$ to $2.50$, repository) the CT has $1.86$ to $2.45$ times the power of pooled $X^2$ at $\varepsilon=0.06$ ($1.16$ to $1.70$ at $0.10$) and $2.1$ to $5.7$ times that of unpooled $X^2$, the unpooled gap largest where many bins expect fewer than one count (cell, moon). AC reaches $0.72$ to $0.79$, its Tippett combination with $X^2$, $0.64$ to $0.70$: with the period known to be $2$ these are preferable; against period-4 code widths on the same five images both fall to $0.14$ or less, while the CT keeps $0.14$ to $0.56$, ahead of pooled $X^2$ on two (repository). Block 3 is the IEEE~Std~1241 sine-wave test~\cite{ieee1241}, arcsine null, random-phase sampling; gamma calibration gives the CT $0.770$, MC $0.772$. With amplitude and offset fitted from the same record, the precomputed cut makes the CT conservative (size $0.025$, power $0.65$, $X^2$ $0.44$), and a parametric bootstrap cut gives size $0.048$ and power $0.74$ (repository).

\begin{table*}[!t]
\centering
\caption{(iii), (iv): Power at $\alpha=0.05$, MC-calibrated}
\label{tab:shapes}
\scriptsize
\setlength{\tabcolsep}{4pt}\renewcommand{\arraystretch}{0.92}
\begin{tabular}{@{}llc cccccccccc@{}}
\toprule
null & kind & index & CT & $T(1)$ & min-$p$ & $X^2$ & $X^2_p$ & CvM & AD & $d_{\mathbf u}$ & AC & vN \\
\midrule
tent $1{:}2{:}3{:}4{:}3{:}2{:}1$ & add. & 0.99 & 0.424 & \textbf{0.485} & 0.468 & 0.414 & -- & 0.063 & 0.082 & 0.002 & 0.588 & 0.212 \\
tent $1{:}2{:}3{:}4{:}3{:}2{:}1$ & mult. & 0.99 & 0.499 & 0.498 & \textbf{0.500} & 0.329 & -- & 0.067 & 0.069 & 0.003 & 0.590 & 0.353 \\
run-length $16{:}8{:}4{:}2{:}1{:}1$ & add. & 0.96 & 0.618 & \textbf{0.842} & 0.816 & 0.806 & -- & 0.108 & 0.119 & 0.251 & 0.641 & 0.280 \\
run-length $16{:}8{:}4{:}2{:}1{:}1$ & mult. & 0.96 & \textbf{0.573} & 0.490 & 0.547 & 0.382 & -- & 0.339 & 0.309 & 0.002 & 0.637 & 0.119 \\
\midrule
coins & width & 0.16 & \textbf{0.317} & 0.242 & 0.293 & 0.106 & 0.168 & -- & -- & 0.278 & 0.774 & 0.267 \\
cell & width & 0.90 & \textbf{0.663} & 0.222 & 0.591 & 0.132 & 0.271 & -- & -- & 0.001 & 0.787 & 0.402 \\
moon & width & 0.95 & \textbf{0.650} & 0.209 & 0.591 & 0.115 & 0.317 & -- & -- & 0.000 & 0.749 & 0.407 \\
\midrule
converter, sine & width & 0.42 & \textbf{0.772} & 0.762 & 0.768 & 0.508 & -- & -- & -- & 0.641 & 0.991 & 0.535 \\
\bottomrule
\end{tabular}

\vspace{2pt}\parbox{\textwidth}{\scriptsize Standard error at most $0.005$; higher is better; bold: best except AC. Block 1: period-2 comb at $\ell_1$ distance $0.10$, $N=500$, additive or multiplicative (Section~\ref{subsec:family}). Block 2: three of the five image code histograms, $n=64$, $N=2000$, code widths $\propto1+(-1)^{i+1}\varepsilon$, $\varepsilon=\pm0.06$. Block 3: IEEE~Std~1241 sine-wave test, arcsine null, $n=64$, $N=10^4$, mismatch $\varepsilon=\pm0.05$. Entries: per statistic, the worse comb sign; index: collapse index at that $N$. $X^2_p$: bins pooled to expected count five (--: no pooling needed, or not run). CT $=T(0)$, min-$p$, $T(1)$: Section~\ref{subsec:family}; $\du$: uncentered, one-sided; AC: alternating contrast $\A{\sum_{i=1}^{n}(-1)^i e_i}$, blind to the comb sign but tied to period~2; vN: von Neumann ratio $\sum_{i=1}^{n-1}(e_{i+1}-e_i)^2/\sum_{i=1}^{n}e_i^2$.}
\end{table*}

\subsection{Application: run-length testing of a bit source}
\label{subsubsec:runs}

For a fair bit source (independent, equiprobable bits) run lengths are independent, with $\Pr[\text{run}=k]=2^{-k}$; pooling runs of $6$ or more gives $\mathbf{q}=(16,8,4,2,1,1)$. The defect is two interleaved lanes (position parities, unknown to the tester) of different flip rates, $\Pr[b_\tau\neq b_{\tau-1}]=1/2+(-1)^\tau\varepsilon$ for bit $b_\tau$ at position $\tau$, like an interleaved converter~\cite{kurosawa2001explicit}; the mismatch is smooth: bit frequency stays balanced while even run lengths gain at odd's expense, a period-2 comb on a steep geometric null. The effect is second order, $\Pr[\text{run}=1]$ falling by exactly $2\varepsilon^2$: $\varepsilon=0.10$ is an $\ell_1$ departure of $0.07$. The source is simulated; an odd run swaps the lane, so run lengths are dependent; their marginal law is measured from $4\cdot10^7$ bits.

With $N=200$ runs the exact rule $S>1560$ has size $0.0485$; on histograms drawn from the source's run-length law (MC, $10^5$ draws) the CT has power $0.169$ at $\varepsilon=0.10$ and $0.603$ at $\varepsilon=0.15$, $X^2$: $0.103$ and $0.409$. Fig.~\ref{fig:main}c ($2000$-bit windows, $\approx1000$ runs) compares both with six NIST tests~\cite{rukhin2010statistical}, MC-calibrated on the same windows at block lengths $m$ favoring them. Frequency, runs, spectral, and serial at $m=2$ stay within $0.024$--$0.076$ in every window, whether the starting lane is fixed or drawn (repository). The nearest NIST test, longest run, is built on run lengths: it leads in the wider windows at the smallest amplitudes, by up to $0.19$; the CT leads in $400$-bit windows throughout and is at or above it from $\varepsilon=0.10$. Drawing the starting lane per window, as a capture would, weakens that test wherever it has power ($0.548$ down to $0.308$ at $\varepsilon=0.10$), the CT not at all. In these windows the uncentered $\du$ would fall to $0.011$ at $\varepsilon=0.10$: the collapse of Section~\ref{sec:degeneracy} on applied data. Tests aimed at the lanes, a serial test on non-overlapping bit pairs or an even/odd split of the flips, detect the defect far earlier, $0.76$ and $0.99$ against the CT's $0.09$ at $\varepsilon=0.05$ (2000-bit windows, random lane, repository): the CT is for defects of unknown structure.

\section{Conclusions}
\label{sec:conclusions}

For a non-uniform null the comb test is defined on centered counts, through the TVC $\dr$; it keeps phase-blindness and one-sidedness and has an exact null for any rational $\mathbf{r}$, to the authors' knowledge a first for the DTV beyond the uniform null, cheap for few bins. The uncentered $\du$ would not be a comb test on a steep null: it would lock into a signed sum of counts, detecting one comb sign and missing the other. With image histograms as nulls the CT has about twice the power of pooled $X^2$ at a 6\% mismatch, but loses to $X^2$ for an additive comb on the run-length null ($0.618$ against $0.806$). The null must be known: one estimated from a clean capture of the test sample's length inflates the size of the CT and of $X^2$ alike, which a $125$ times longer capture removes for the CT (repository); it is known analytically for the bit source and, up to amplitude and offset that a bootstrap cut absorbs, for the converter. Future work includes exact finite-$N$ null moments, under development by the authors, to remove MC from calibration; the two-sided variant of~\cite{banic2026comb}; the circular DTV; minimax rates of comb detection~\cite{ingster2003nonparametric,kipnis2026minimax}; and double-quantization combs~\cite{popescu2004statistical,lukas2003estimation}.

\section*{Acknowledgment}
Generative artificial intelligence (Claude Opus~5 and~5.5, Fable~5.1; Anthropic) helped with English and editing, drafted the code, checked the authors' derivations, and read the Russian source~\cite{zubkov2002computational}; the authors verified code and numbers and take responsibility for all content.

\IEEEtriggeratref{17}
\bibliographystyle{IEEEtran}
\bibliography{literature}

\end{document}